\documentclass[intlimits,twoside,a4paper]{article}
\usepackage{amsmath,amssymb}
\usepackage{amsfonts,bm,amssymb}
\usepackage[T2A]{fontenc}
\usepackage[cp1251]{inputenc}

\usepackage[eqsecnum]{cmpj2}

\usepackage{graphicx}
\usepackage{amssymb}
\usepackage{amsmath}
\usepackage{epstopdf}
\usepackage{color}
\usepackage{soul}
\usepackage{ulem}
\usepackage[textsize=tiny]{todonotes}
\usepackage[12pt]{moresize}

\usepackage{breakurl}

\RequirePackage{color}

 \definecolor{MyDarkGreen}{rgb}{0.02,0.60,0.06}

\issue{2017}{20}{1}{13803}
\doinumber{10.5488/CMP.20.13803}

\title[A scientists' view of scientometrics:  Not everything that counts can be counted]
{A scientists' view of scientometrics:  Not everything that counts can be counted\footnote{The phrase ``not everything that can be counted counts, and not everything that counts can be counted'' comes from William Bruce Cameron's 1963 text {\textit{Informal Sociology: A Casual Introduction to Sociological Thinking}}. It is frequently ascribed to Albert Einstein but that link is not solidly supported \cite{QI}.}}

\author[R.~Kenna, O.~Mryglod, B. Berche]{R.~Kenna\refaddr{ad1,ad4}, O.~Mryglod\refaddr{ad2,ad4}, B. Berche\refaddr{ad3,ad4}}
\addresses{
\addr{ad1}
Applied Mathematics Research Centre, Coventry University, CV1 5FB, England
\addr{ad2}
Institute for Condensed Matter Physics of the National Academy of Sciences of Ukraine, \\
1 Svientsitskii St., 79011 Lviv, Ukraine
\addr{ad3}
Statistical Physics Group, Universit\'e de Lorraine, IJL, UMR CNRS 7198,
Campus de Nancy,\\ B.P. 70239, 54506 Vand\oe uvre l\`es Nancy Cedex, France
 \addr{ad4} The Doctoral College for the Statistical Physics of Complex Systems,
Leipzig-Lorraine-Lviv-Coventry $({\mathbb L}^4)$, Europe
}

\authorcopyright{R.~Kenna, O.~Mryglod, B. Berche, 2017}
\date{Received January 19, 2017, in final form March 3, 2017}

\begin{document}

\maketitle
\begin{abstract}
Like it or not, attempts to evaluate and monitor the quality of academic research have become increasingly prevalent worldwide.
Performance reviews range from at the level of individuals, through research groups and departments, to entire universities.
Many of these are informed by, or functions of, simple scientometric indicators and the results of such exercises impact onto careers, funding and prestige.
However, there is {sometimes} a failure to appreciate that scientometrics are, at best, very blunt instruments {and their incorrect usage {can be misleading}}.
Rather than accepting the rise and fall of individuals {and institutions} on the basis
of such imprecise measures, {calls have been made for indicators be regularly scrutinised
and for improvements to  the evidence base in this area.}
It is {thus} incumbent upon the scientific community, especially the physics, complexity-science {and scientometrics} communities, to  {scrutinise} {metric indicators.}
Here, we review recent attempts to do this and show that some metrics in widespread use  {cannot be used} as reliable indicators research quality.

\pacs 89.20.-a, 89.65.-s
\end{abstract}

\vspace{1mm}

This paper is dedicated to Professor Yurij Holovatch on the occasion of his 60th birthday and in recognition of his many important contributions to statistical physics, including sociophysics and related areas.

\section{Introduction}
\vspace{1mm}

The field of scientometrics can be traced back to the work of the physicist Derek de Solla Price~\cite{Pr65} and the {linguist/businessman} Eugene Garfield~\cite{Ga79}.
It is the quantitative study of the impact of science, technology, and innovation~\cite{LeMi15}.
This frequently involves analyses of  citations and facilitates, (indeed, encourages) the evaluation and ranking of individual scientists, research groups,  universities and journals.
The closely related (sub-)field of bibliometrics is concerned with measuring the impact of scholarly publications.
Perhaps the most famous indicator of the productivity and impact of a scientist is the so-called $h$-index (named by its creator Jeorge E. Hirsch) \cite{Hi05}
and the most famous indicator of journal quality is the impact factor, devised by Garfield and Irving Sher \cite{Ga55,Ga05}.

The UK is at the forefront of group or departmental research evaluation and has been for a number of decades.
The first Research Assessment Exercise (RAE) was carried out in 1986, introducing an explicit, formalised assessment process of research quality.
The RAE was adapted and developed to a more comprehensive process and subsequent exercises were undertaken in 1989, 1992, 1996, 2001 and 2008.
The exercises were not based on citation counts and concerned the research quality of whole units or research groups that are put forward for assessment, rather than individuals.
Panels of experts were assembled from different academic disciplines and used to evaluate the quality of research taking place across the UK academy.
The Research Excellence Framework (REF) was introduced as the successor to the RAE and was undertaken in 2014 to assess the research carried out between 2008 and 2013. This exercise was also based on peer review by panels of subject experts.
{Thus, peer-review-based assessment exercises have been in use in the UK for over 30 years and, although far from perfect \cite{Sayer}, are widely seen by the academic community as the only acceptable approach currently in existence.
For example, a recent, influential, national review, which received 153 responses to its call for evidence from interested parties, found that  ``a common theme that emerged was that peer review should be retained as the primary mechanism for evaluating research quality''
and that peer review ``should continue to be the ``gold standard'' for research assessment'' \cite{Tide1}.}

The issue of whether or not scientometrics and bibliometrics should be used for the RAE or REF {and, indeed, other national exercises,} is one that has been continuously debated.
Some suggest that metrics should form an integral part of such exercises or, indeed, replace them entirely.
Others completely reject this idea and advocate that only peer review can be trusted.
A middle way has also been suggested --- that metrics be used to inform assessors in some subject areas.
{(There are regular discussions in national press on such matters, see, e.g., \cite{Sayer, Bi14, Bi17,Jump,caution,Jump2}.)}

Here, we report on our analyses which compared the outcomes of scientometrical measurements of research with those coming from the RAE/REF peer-review systems.
We consider two metric indicators:  the $h$-index \cite{Hi05} and the so-called normalised citation index (NCI)~\cite{Ev10,Evidence2011}.
First, we summarise the outcome of studies which demonstrate the role and importance of group size in research evaluation
\cite{KeBe10,KeBe11a}.
Then, we demonstrate that the NCI is a poor proxy for peer-review measures of research quality and, although the NCI is well correlated with research strength {of large groups} in some disciplines \cite{MrKe13a}, the departmental $h$-index is slightly better \cite{MrKe13b}.
We then report on an attempt to use the latter to predict outcomes of REF2014 \cite{MrKe15a} and show that these predictions
were wildly off the mark \cite{MrKe15b}.
Our conclusion is that metrics, at least in their current form, should not be used as proxies for measures of research quality.

\section{The RAE and REF; why size matters; the $h$-index and the NCI}

For the RAE in 2008, three aspects of group or departmental quality were considered: research outputs, research environment and research esteem.
The first of these mostly entailed publications, but for some disciplines software, patents, artefacts, performances or exhibitions were also  considered.
Research environment was also quantified at RAE2008  and institutions were asked to provide information on funding, infrastructure, vitality, leadership, training, accommodation and so on.
The third component for RAE2008 was esteem and indicators included prizes, honours, professional services and other activities.
The precise manner in which  outputs,  environment and esteem fed into the overall final RAE score was dependent upon
discipline; in pure and applied mathematics, statistics and the computer sciences, they were weighted at 70\%, 20\% and
10\%, respectively, while in biology and other subjects they were weighted at 75\%, 20\% and 5\%, respectively.

RAE2008 estimated the research quality of each submitted research unit in a number of academic disciplines.
These estimates were presented as profiles, detailing the proportions of research activity carried out at each of five levels:
4* (world-leading research);
3* (internationally excellent research);
2* (research that is internationally recognised);
1* (research recognised at a national level) and unclassified research.
Following the exercise, a formula was used to determine how funding is allocated to higher education institutes for the subsequent years.
The formula used by the Higher Education Funding Council for England, immediately after {RAE2008,} valued 4* and 3* research seven and three times, respectively, more than 2* research and allocated no funding to 1* and unclassified research.
We use that formula to condense the research profile of a unit into a scalar as follows: if $p_{n\text{*}}$ represents the percentage of  a team's research which was rated  $n$*, then a proxy for the team's \textit{quality} is
\begin{equation}
 s = p_{4\text{*}} + \frac{3}{7}p_{3\text{*}} + \frac{1}{7}p_{2\text{*}}
\,.
\label{seven}
\end{equation}
The research ``strength'' of a unit\footnote{Research strength as defined above can be compared to the so-called ``research power'', which is a measure that has recently gained in popularity in the UK.
Research power is the simple grade point average of a submission ($= 4\text{*}p_{4\text{*}} + 3\text{*}p_{3\text{*}} + 2\text{*}p_{2\text{*}} + p_{1\text{*}}$) multiplied by $N$.
Other measures are possible; following lobbying by pressure groups the funding formula (\ref{seven}) changed a number of times to concentrate more money into those groups with the highest quality profiles. Here, we stick with formula (\ref{seven}) as it has the advantage of clearly demarcating four quality levels prior to political influence. We have checked that small changes in the formula do not deliver changes to the outcomes of our analysis.
} is then given by $S=sN$, where $N$ is the size of the submitted team.
The amount of money flowing into the university from the Higher Education Funding Council for England was then a function of $S$.

When research quality $s$ is plotted against  $N$, an interesting pattern emerges for many subject areas.
It was shown in \cite{KeBe10,KeBe11a} that quality increases linearly with size up to a certain point, identified as the discipline-dependent point at which research groups tend to become unwieldy and may start to fragment.
This is similar to the {\textit{Ringelmann effect}} \cite{Ri13} in social psychology and marked by a {\textit{Dunbar number}} \cite{Du92} which is discipline dependent.
A statistical-physics-inspired, mean-field-type theory exposes the existence of a second important group size which may be identified as the critical mass and is also dependent on discipline \cite{KeBe10}.
For theoretical physics, for example, the critical mass is 6.5 and the Dunbar number is~13. For experimental physics, the corresponding numbers are 13 and 25, respectively.
With the critical mass and Dunbar numbers to hand, one may classify groups according to their size.
Small groups are below critical mass; medium ones are bigger than critical mass but smaller than the Dunbar number;
 and groups of still more members are classified as large.
Thus, a group of 15 theoretical physicists would be deemed large, for example, while the same number of experimentalists would be considered as medium in size.
We will shortly see that size matters when comparing metric indicators to peer-review estimates of research quality.

For REF2014, a number of changes were introduced vis-\`a-vis  RAE2008.
Firstly, that the esteem category was replaced by impact.
The latter, not to be confused with academic or citation impact, was defined as ``an effect on, change or benefit to the economy, society, culture, public policy or services, health, the environment or quality of life, beyond academia'' \cite{HEFCE2012a}.
The three categories outputs, environment and impact were then weighted at 65\%, 15\% and 20\%, respectively.
Another change was that, while for the RAE research was categorised into 67 academic disciplines, in the 2014 REF there were only 36 units of assessment.
The Applied Mathematics {Unit of Assessment}, for example, (which included some theoretical physics groups), was a category at RAE2008, but for REF2014 it was merged with Pure Mathematics, Statistics and Operational Research.
One may argue, therefore, that RAE2008 was more ``fine-grained'' than REF2014.

The next REF is expected to take place in 2021.
The rules have not yet been decided, but it is expected that it will build upon REF2014 although there will be incremental changes.
The precise role of metrics at REF2021 is yet to be decided but indications are that peer review should remain the primary method of research assessment.
Our analysis strongly supports this direction --- not only for the UK, but for all national exercises of this type.

The question we wish to address is whether or not metrics such as the $h$-index or NCI should be used for exercises such as the REF.
The $h$-index seeks to measure the citation impact of a researcher along with the volume of their productivity.
It is defined as the number of papers an author has produced that each have been cited $h$ times or more.
Its scalar simplicity renders it very attractive to policy makers and managers.
Although originally introduced as a measure at an individual level, the $h$-index can also be applied  to estimate the productivity and scholarly impact journals, research  groups, departments or  universities \cite{JoHu11,Bi14}.

{\textit{Thomson Reuters Research Analytics}} has developed the so-called {\textit{normalised citation impact}} (NCI) as another
measure of a department's citation performance in a given discipline \cite{Ev10,Evidence2011}.
A useful feature is that it attempts to take account of differences in citation rates across different  disciplines by ``rebasing'' the total citation count for each paper to an average number of citations per paper for the year of publication and either the field or journal in which the paper was published.
The measure is determined for an entire group or department and then normalised by the group size.
It is, therefore, a {\textit{specific}} (per-head) measure (also called {\textit{intensive}} in the parlance of statistical physics).
Scaled up to the size of a group or department, the corresponding {\textit{absolute}} measure  ({\textit{extensive}}  in statistical-physics terminology).
Here, we denote the specific NCI  by $i$ and its absolute counterpart by $I$ where $I = iN$.

In section~\ref{Can}, we report on a quantitative comparison of both of these indicators against expert peer review measures of the quality of research groups coming from RAE/REF after a brief qualitative discussion in section~\ref{Should}.

\section{Should metrics be used in the research evaluation schemes?}
\label{Should}

The debate as to whether metrics should or should not be used in national evaluation frameworks is a long one within the academic, scientometric,
university-management and policy-making communities internationally.
Although flawed in many ways, systems based purely upon peer review enjoy the highest confidence of the scientific community itself {\cite{Tide1}}.

Flaws include the absence of trusted methods to account for different levels of expertise, stringency and bias amongst assessors and the absence of an acceptable way to normalise results across different disciplines.
(A new approach to overcome some such difficulties  has recently been developed \cite{MaPa17}.)
Another objection is that peer-review-based exercises such as the RAE and REF are also expensive~\cite{cost1}.
It has been estimated that the total cost to the UK of running REF2014 was \pounds246M.
That amount comprises \pounds14M in costs for the UK higher-education funding bodies which run the exercise and \pounds232M in costs to the higher education community itself.
The latter figure includes \pounds19M for the panellists' time and \pounds212M for preparing the REF submissions (about \pounds4K for each of the 52\,077 researchers  submitted).
Costs are, therefore, a prime reason forwarded by advocates for replacing peer-review exercises by automated systems based on metrics.
Another is the burden in terms of time taken away from research activity in order to prepare REF submissions.\footnote{
We are reminded of the novel {\it The Mark Gable Foundation} by Leo Szilard, in which advice  to {\textit{retard}} scientific progress is:
``Take the most active scientists out of the laboratory and make them members of \dots committees.
And the very best \dots should be appointed as Chairmen''.
In this way ``the best scientists would be removed from their laboratories and kept busy on committees passing on applications for funds.
Secondly,  the scientific workers in need of funds will concentrate on problems which are considered
promising and are pretty certain to lead to publishable results. \dots By going after the obvious, pretty soon Science will dry out.
Science will become something like a parlor game. Some things will be considered interesting, others will not. There will be
fashions. Those who follow the fashion will get grants. Those who won't, will not,
and pretty soon they will learn to follow the fashion too''~\cite{Szilard}.
}

Metrics were not officially used in the earlier RAE or in the REF, although there was nothing to prevent individual assessors from determining the citation counts of
individual papers or looking up the citation records of individual researchers.
For the 2014 exercise, REF panels were allowed to use citation data, but only to inform their judgements (e.g., to decide how academically significant a paper was) which were predominantly based on peer review (assessors were advised to recognise the significance of outputs beyond academia as well).
To this end, citation data were sourced centrally by the REF team using the Scopus database.
Assessors were, however, instructed not to refer to additional bibliometric data, such as impact factors or other journal-level metrics in their deliberations.

{In comparison, in} France, prior to the creation of the AERES (Agence d'\'Evaluation de la Recherche et de l'Enseignement Sup\'erieur) in 2006, research assessment was essentially performed by the CNRS (Centre National de la Recherche Scientifique) solely on the basis of evaluation by the peers.
Panels of peers were composed in assessment committees, visiting the laboratories  they were {assigned to}.
In a given discipline, there were many different panels of experts of this kind.
This format of a committee of pairs visiting the laboratories remained with the AERES and then with  its successor, the HCERES (Haut Conseil de l'\'Evaluation de la Recherche et de l'Enseignement Sup\'erieur) since 2013.
The novelty introduced with the AERES is the scale of evaluation campaigns and the field of expertise, evaluation being performed at the scale of research teams as well as the scale of universities, or even evaluation of the CNRS itself!
 Of course, the use of bibliometrics was progressively introduced into the reports and as a guiding element for the evaluation. AERES had even marked the laboratories and universities according to a rating system A$^+$, A, B, and C, similar to that of the British system. This rating system is now abandoned.

The Australian Research Council used Scopus as the citation and bibliometrics provider for the Excellence in Research for Australia (ERA) schemes both in 2010 and 2012.
Italy's Research Evaluation Exercise  will use ``informed'' peer review.
This means that, in areas such as  the mathematical, natural, engineering and life sciences, peer evaluation will be supported by bibliometric information from the Web of Science and Scopus citation databases.
Evaluators will use both information about the impact of individual articles (through numbers of citations) and the quality of the journals in which they are published (through the Impact Factor and other indicators).
In humanities and the social sciences, however, the system uses peer evaluation only.

There is no single procedure to assess research institutions in Ukraine. Regular evaluations are mostly based on formal
reports, supported by scientometrics. However, their use is often rather {haphazard.
Dangerously attractive, simple metrics are sometimes} used without clear understanding of their peculiarities.

Many scientists and other academics object to the {misuse of the} scientometric quantification of their research.
A fundamental objection is that the metrics are doomed to fail {if} their intended task {is to aid} management and funding of science by making it  systematic and objective.
{In 1977, Garfield himself cautioned against the misuse of citation analyses \cite{caution}.
In the forty years since, however, those words appear to have fallen on deaf ears as citation-misuse is rife  \cite{Tide1}.
{In response, the {\textit{San Francisco Declaration on Research Assessment}} \cite{DORA} was initiated by a group of experts, editors and publishers to call for improvements in the ways in which scientific research is evaluated.}
{Similarly motivated by the fact that research evaluation ``is increasingly driven by data and not by expert judgement'',
the {\textit{Leiden Manifesto for Research Metrics}} has been drawn up, comprising
 ten principles for the measurement of research performance \cite{Leiden}.
{In the UK,} the ``Metric Tide'' steering group \cite{Tide1} felt it necessary to  set up a website as a forum for ongoing discussion of these issues; to ``celebrate and encourage responsible uses of metrics'' but also to ``name and shame bad practices when they occur''. (Every year they plan to award a ``Bad Metric'' prize to the most inappropriate use of quantitative indicators \cite{bad}.)

{It} is claimed that {the misuse of} such metrics  {is} changing the nature of science; they are damaging curiosity-driven research as scientists are forced to maximise their personal metrics instead.
In a system which excessively rewards novel findings over confirmatory studies, the most rational research strategy is for scientists to spend most of their effort seeking novel results through small studies with low statistical power \cite{HiMu16}.
As a result, half of the studies they publish would contain erroneous conclusions.
The existence of a ``trade-off'' between productivity and rigour was also claimed in \cite{SmMc16}:
poor methods result from incentives that favour them and one of these is the priority of publication over discovery for career advancement.
These are examples of Goodhart's law: when a quantitative metric is introduced as a proxy to reward academics, these metrics become targets and cease to be good measures \cite{Goodhart,Ball2016}.

Notwithstanding these objections, we next ask whether or not metrics are  capable of approximating the results of RAE or REF.
{Again, our task is motivated by the widespread view of peer review as a ``gold standard''
\cite{Tide1} and the desires by some to replace or inform it. }
We shall find that at least the NCI and $h$-index are not {capable of approximating RAE/REF}.
This suggests that the UK should persist with its peer-review based REF-type evaluation system and that other countries should also seek to move in this direction and away from metrics-driven exercises.

\section{Can metrics be used as a proxy for peer review?}
\label{Can}
\subsection{The NCI}
\begin{table}[!t]
\caption{Correlation coefficients between absolute values $I$ and $S_1$ (boldface) and specific values $i$ and $s_1$ (regular typeface) calculated for several different disciplines. Here, the subscript 1 indicates that only the ``outputs'' sector of the RAE results are used. Pearson's  correlation coefficient $r$ is presented for all groups in a given discipline and separately for large groups and small/medium groups. We also present Spearman's coefficient for ranked values for all groups.}
\vspace{2ex}
\begin{center}
\begin{tabular}{|l|l|l|l|c|}
\hline\hline
                             & \multicolumn{3}{|c|}{Pearson coefficient $r$}& Spearman \\
\cline{2-4}
Discipline                   &for     &for    &for           & coefficient   \\
                             &all     &large  &medium/small  & of ranked      \\
                             &groups  &groups &groups        & values $\rho$  \\
\hline
\hline
Biology                      &{\bf{0.97}}~~0.60    &{\bf{0.96}}~~0.57   &{\bf{0.90}}~~0.35    & 0.53   \\\hline
Chemistry                    &{\bf{0.96}}~~0.60    &{\bf{0.96}}~~0.82   &{\bf{0.79}}~~0.34    & 0.62   \\\hline
Physics                      &{\bf{0.96}}~~0.48    &{\bf{0.96}}~~0.45   &{\bf{0.67}}~~0.54    & 0.53   \\\hline
Engineering                  &{\bf{0.92}}~~0.34    &                    &                     & 0.18   \\\hline
Geography                    &{\bf{0.88}}~~0.51    &{\bf{0.56}}~~0.13   &{\bf{0.93}}~~0.42    & 0.47   \\\hline
Sociology                    &{\bf{0.88}}~~0.49    &{\bf{0.82}}~~0.29   &{\bf{0.73}}~~0.64    & 0.47   \\\hline
History                      &{\bf{0.88}}~~0.34    &{\bf{0.79}}~~<0     &{\bf{0.66}}~~0.27    & 0.38   \\
 \hline\hline
\end{tabular}
\label{tab1}
\end{center}
\end{table}
\vspace{-3mm}
\begin{figure}[!t]
\begin{center}
\includegraphics[width=0.44\columnwidth,angle=0]{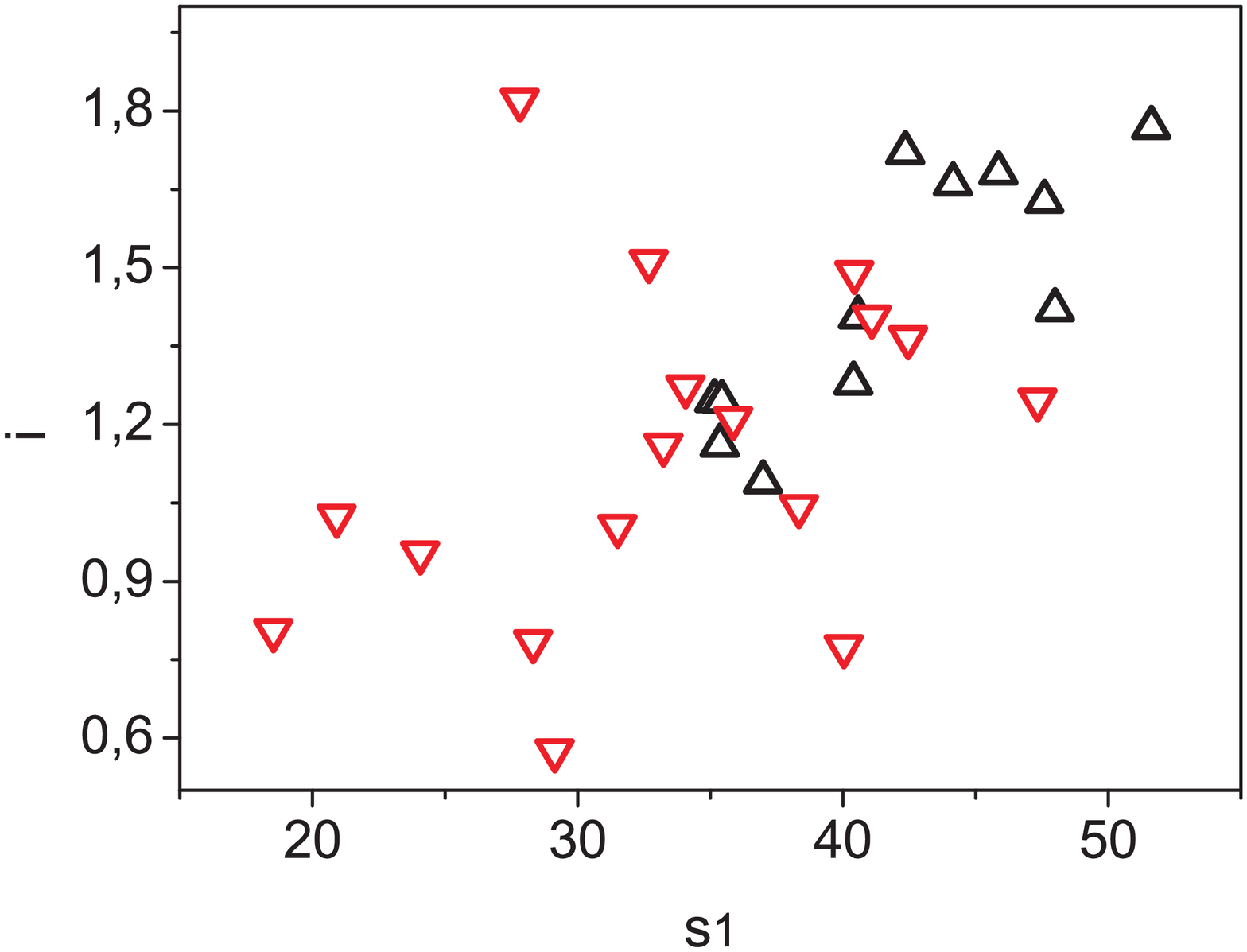}
\hspace{5mm}
\includegraphics[width=0.44\columnwidth,angle=0]{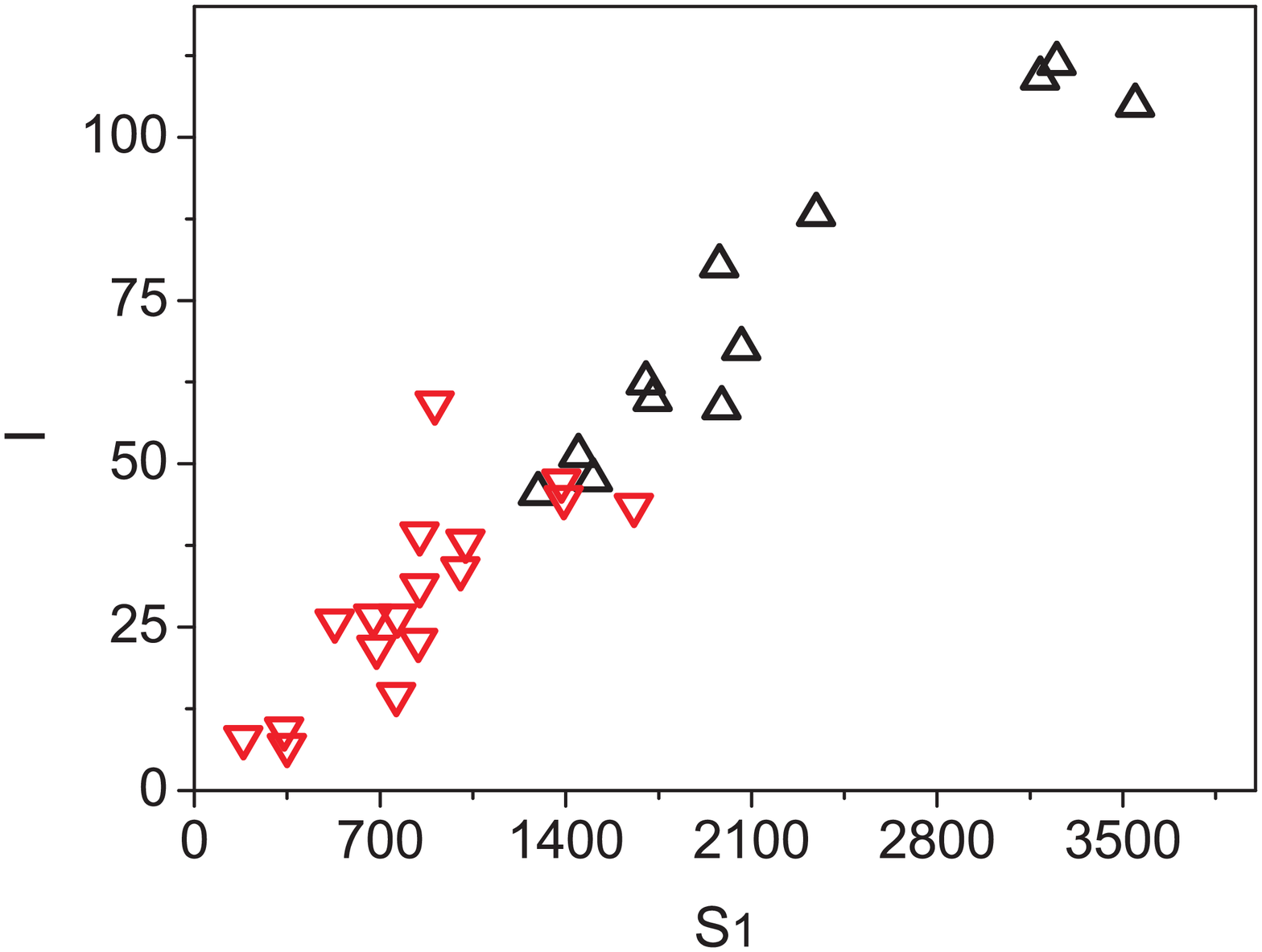}
\end{center}
\vspace{-3mm}
\caption{(Color online) The left-hand panel depicts the correlations between the normalised citation impact $i$ and RAE peer-review measures of group research quality $s_1$ for the discipline of chemistry.
The right-hand panel gives the correlations between the absolute  impact $I=iN$ and  research strength $S_1=s_1 N$ for the same discipline.
The black $\triangle$ symbols represent large research groups and the  red  {$\bigtriangledown$} symbols represent   medium/small groups.}
\label{fig2}
\end{figure}

In \cite{MrKe13a,MrKe13b}, NCI values were compared with RAE2008 measures of research quality and strength for various  groups in various disciplines,  from the natural to social sciences and humanities.
The results are reproduced here in table~\ref{tab1}.
(We refer the reader to the original literature for tests of significance~\cite{MrKe13a,MrKe13b}.)
Actually,  only the outputs component of RAE results are used in the determination of the correlation coefficients in this instance --- i.e., neither the environment nor the esteem measures are used here.
This is because only outputs contribute directly to the NCI.
We label the corresponding quality and strength measures by $s_1$ and $S_1$, respectively.
The table lists the values of the Pearson correlation coefficient for extensive (absolute) quantities (namely $I$ vs. $S_1$) in boldface and those for intensive (specific) quantities (namely $i$ vs. $s_1$) are given in regular typeface.
Figure~\ref{fig2} gives examples of $i$ vs. $s_1$ and $I$ vs. $S_1$ plots for the case of chemistry research groups.

One observes that the  intensive measure $i=I/N$ is poorly correlated with group quality $s_1 = S_1/N$ for all disciplines and for all group sizes.
One of the best correlations between $i$- and $s_1$-values is in the case of chemistry, but even then the Pearson correlation coefficient is only 0.6.
Since NCI and RAE scores are also used to rank research groups, we also evaluated the Spearman correlation coefficient between such rankings.
The highest value is for chemistry and that is only 0.62.
This means that the NCI fails as a proxy for RAE measures of group research quality and also fails to deliver a ranking anywhere similar to that delivered by the RAE.

Poor correlations are evident for chemistry in the left-hand panel of figure~\ref{fig2}.
On the other hand, the absolute indicator $I$ is very highly correlated with the peer-evaluated measures for physics, chemistry and biology where the Pearson correlation coefficient is 0.96 or above.
The right-hand panel of the figure exhibits better correlations as multiplication of intensive quantities by $N$ ``stretches'' the data along each axis.
On closer examination, however, these correlations are the best for large groups in these disciplines; for small/medium groups, they fall below 0.90.
The correlations are also worse  for other disciplines, even for their large groups; e.g., for sociology and history they were 0.88.

This outcome suggests the almost paradoxical result that the NCI could possibly form a basis for deciding on funding amounts for research institutions, but only for the sciences and only for large research groups. It should not be used in any other cases (not for social sciences, humanities and not even for science groups with sub-Dunbar numbers of staff). And certainly, it should never be used as a basis for ranking or comparing the research groups. Further details are given in \cite{MrKe13a,MrKe13b}.

\subsection{The departmental $h$-index}

At this stage, we have established that the NCI is not a good specific (intensive) indicator for research quality.
Is there a better metric, perhaps?
In \cite{MrKe15a}, we demonstrated that the {\textit{departmental $h$-index}} \cite{JoHu11,Bi14} has indeed a better correlation with the RAE-measured strength index $s$ than has the NCI, $i$.
A departmental $h$-index of $n$, say, indicates that $n$ papers authored by
researchers from a given department  in a given discipline  were cited at least $n$ times over a
given time period.
The departmental $h$-index uses data from {\textit{all}} researchers from a given department, not only those submitted to RAE or REF.
However, in practice, individuals with weaker citation records are swamped out by those with stronger ones, so that it can be dominated by a few individuals --- even by a single, extremely strong one.

We determined departmental $h$-indices for universities which submitted to RAE2008 within the
disciplines of biology, chemistry, physics and sociology.
The citation data we used were taken from the Scopus database.
In order to estimate $h$, we filtered the Scopus data to extract only those publications which correspond to UK and which were published in the period 2001--2007, so to compare with RAE2008.
We selected  subjects most closely corresponding to the above four disciplines using Scopus subject
categories (see \cite{MrKe15a} for details).
Unlike for the RAE and REF, where authors' affiliations  are determined by their addresses at the assessment census date, the author address at the time of publication determines to which university a given output is allocated for the departmental $h$-index.
A small number of institutions were not listed in the Scopus database after refining the search result, so  it was not possible to determine $h$-indices in these cases.
This means that the set of universities contributing to table~\ref{tab2} slightly differs from that contributing to table~\ref{tab1}.
The results we present in table~\ref{tab2} are for institutions that could be accessed by Scopus.
The table shows that the departmental $h$-index is indeed better correlated with overall RAE-measured research quality.
However, the correlations between the $h$-index and the RAE results are still too small to replace the peer-review exercise by metrics.
{(We again refer the reader to the original literature for significance tests~\cite{MrKe13a,MrKe13b}.)}

\vspace{-2mm}
\begin{table}[!h]
\caption{Correlation coefficients between metrics and  RAE2008 measures of  research quality.
The first column compares the departmental $h$-index measured at the beginning of 2008 with the RAE quality scores for research outputs. The second column suggests that there is a better correlation between $h$ and the overall RAE scores. Comparing to the third column, one sees that the departmental $h$-index delivers a better correlation than the NCI.}
\vspace{2ex}
\begin{center}
\begin{tabular}{|l|l|l|l|}
\hline\hline
                             & \multicolumn{3}{|c|}{Pearson coefficient $r$}                \\ \cline{2-4}
Discipline                   &$h_{\rm{2008}}$ vs. $s_1$ &$h_{\rm{2008}}$ vs. $s$   &$i$ vs. $s$ \\
\hline
\hline
Biology                      & 0.65                                & 0.74                    & 0.67 \\\hline
Chemistry                    & 0.74                                & 0.80                    & 0.58 \\\hline
Physics                      & 0.44                                & 0.55                    & 0.37 \\\hline
Sociology                    & 0.57                                & 0.62                    & 0.51 \\
 \hline\hline
\end{tabular}
\label{tab2}
\end{center}
\end{table}
\vspace{-6mm}
\subsection{(Mis-)Predicting REF}

We have seen that neither the NCI nor the departmental $h$-index have good enough correlations with RAE results  to contemplate replacing the peer-review exercise by metrics.
To demonstrate that forcefully, we decided to use the departmental $h$-index to predict outcomes of RAE2014.
The idea was that if simple citation-based metrics are ever to be used as some sort of proxy for  peer
review, one would expect them to be able to predict  at least some aspects of the outcomes  of such exercises.
Even a limited success might suggest that a citation metric could serve at least as a ``navigator'' --- to help guide  research institutes as they prepare for the expert exercises. For example, research managers may be interested in whether metrics could indicate whether or not they are likely to move up or down the REF league tables in various subject disciplines.

We placed our predictions for the rankings in biology, chemistry, physics and sociology on the arXiv in November 2014 --- before the REF results were officially announced.
These were subsequently published as \cite{MrKe15a}.
After the  REF, the results were announced in December 2014, we revisited our  study \cite{MrKe15b}.
The correlations between the REF results and the $h$-index predictions are given in table~\ref{tab3}.

\begin{table}[!t]
\caption{The values of Pearson's coefficient $r$ and Spearman's rank correlation coefficient $\rho$ for different disciplines for different pairs of measures.
The upper part of the table uses $s$ values from the overall RAE and  REF results ($s_{\rm{RAE}}$ and $s_{\rm{REF}}$, respectively) while the lower part corresponds to the results for outputs only.
Correlations between the predicted and actual directions of the shift (up or down) in the ranked lists are given in the final columns.}
\vspace{2ex}
\begin{center}
\begin{tabular}{|l||r|r|r|r|r|}
\hline\hline
 OVERALL \hspace{0.8cm} & \multicolumn{2}{|c|}{$s_{\mathrm{RAE}}$ vs. $h_{2008}$}& \multicolumn{2}{|c|}{$s_{\mathrm{REF}}$ vs. ${h}_{2014}$}&
$\uparrow \downarrow$\\
\hline
\hline  & $r$ & $\rho$& $r$ & $\rho$&$r$\\
\hline Biology   &0.55&0.61&0.58&0.63&{$-$0.15}\\
\hline Chemistry   & 0.80&0.83&0.84&0.89&{0.05}\\
\hline Physics   &0.49&0.55&0.55&0.50&0.26\\
\hline Sociology   &0.50&{0.39}&0.59&0.60&{0.18}\\
\hline
\hline
OUTPUTS ONLY & \multicolumn{2}{|c|}{$s_{\mathrm{RAE}}$ vs. $h_{2008}$}& \multicolumn{2}{|c|}{$s_{\mathrm{REF}}$ vs. ${h}_{2014}$}&$\uparrow \downarrow$\\
\hline
\hline  & $r$ & $\rho$& $r$ & $\rho$&$r$\\
\hline Biology   &0.44&0.51&0.40&0.42&{$-$0.33}\\
\hline Chemistry   & 0.74&0.71&0.71&0.72&{0.20}\\
\hline Physics   &0.44&0.51&0.39&{0.36}&{0.02}\\
\hline Sociology   &0.41&{0.29}&0.71&0.68&{0.06}\\
\hline\hline
\end{tabular} \label{tab3}
\end{center}
\vspace{-4mm}
\end{table}

Note that the sets of universities contributing to the data in table~\ref{tab3} are again different
{from} those {which} contribute to table~\ref{tab2}.
The reason again is that the universities submitting in units of assessment to REF2014 slightly differ from those  submitting to RAE2008.
To compare the like with the like, table~\ref{tab3} only uses universities which submitted in those particular units of assessment in the 2014 exercise.

Our predictions failed to be  in any way useful in  anticipating REF outcomes.
For example, the Pearson correlation coefficient between the REF-measured $s$ value and the departmental $h$-index in physics was only 0.55.
That for outputs alone was even worse, at 0.39.
In table~\ref{tab3}, besides giving the Pearson's coefficient $r$, we also give Spearman's rank correlation coefficient $\rho$ for various disciplines and pairs of measures.
The correlations between rankings are about as weak as the correlations between quality measures.
For example, the Spearman correlation coefficient between $s$ and $h$ in physics was 0.50 and that for outputs only was 0.36.
To illustrate the uselessness of the $h$-index in this context, one submission from a certain university in a certain subject area was ranked 27th according to the departmental $h$-index but actually came in the seventh place according to the REF.
Replacing REF by a metric-based exercise could have been catastrophic for that particular research group.

We also tried to anticipate whether individual institutions would move up or down in the rankings between RAE2008 and REF2014.
The results for the correlations between our predictions and the actual results are also listed in table~\ref{tab3}.
E.g., for physics, chemistry and biology, it  was 0.26,  0.05 and  $-$0.15, respectively.
If we restrict ourselves to outputs only, the correlations are even worse at 0.02, 0.20 and $-$0.33.
As commented in the press later, one would find better estimates of movement in the league tables by tossing dice!
Our results are published as \cite{MrKe15b,BHKM16}.

\section{Discussion}
\label{end}

The vast majority of academics are opposed to the increased use of automated metrics to monitor  research activity.
A concern is that, {because ``inappropriate indicators create perverse incentives'' \cite{Tide1}, inexpert use of} such metrics to simplify the bases for important judgments, decisions and league tables
 surely lead{s} to violations of the age-old and treasured principle of  academic freedom because researchers are forced to chase citation-based metrics rather than {allowed to} follow where their curiosity leads.
Here, we have reported on a series of publications that show that these fears are well founded; metrics  are a poor measure of research quality.
Our advice to those in authority who are attracted to such simple measures is: do not be fooled by their quantitative nature; they are crude at best, and their misuse can damage academic research.

In recent years, the UK has commissioned at least two major reports on the matter of metrics and research evaluation.
The first of these was the {\textit{Wilsdon Report}}, titled {\textit{The Metric Tide: Report of the Independent Review of the Role of Metrics in Research Assessment and Management}}  \cite{Tide1} and concluded that peer review should remain the primary method of research assessment.
These findings were endorsed by the more recent {\textit{Stern Review}} \cite{Stern}.

As pointed out in \cite{Tide1},
{``There are powerful currents whipping up the metric tide''.
However, ``Across the research community, the description, production and consumption of ``metrics'' remains contested and open to misunderstandings''.
 {The tenth Principle of the Leiden Manifesto calls for scrutinisation of indicators \cite{Leiden} and the San Francisco Declaration  urges ``a pressing need to improve the ways in which the output of scientific research is evaluated'' \cite{DORA}.}
For these and other reasons, ``There is a need for more research on research'' \cite{Tide1}. To respond to {such calls}, it is} important that scientists turn their tools to their own discipline too.
Indeed, {amongst other academic evidence,} the {\textit{Metrics Tide}} report \cite{Tide1} made considerable use of \cite{MrKe13b,MrKe15a,MrKe15b}, {which, in turn built upon \cite{MrKe13b,MrKe13a}, which themselves were inspired by statistical-physics mean-field-inspired theories \cite{KeBe10,KeBe11a}.}
{In this paper, we have tried to contextualise such ``research on research'' within the UK national context while highlighting their implications internationally too.}

Moreover, and as pointed out in \cite{cost1}, the \pounds$4\,000$ that it cost to submit and evaluate each  researcher to REF amounts to only 1\% of what it costs to employ them over a six-year period.
Viewed in this way, REF is actually a rather cheap exercise.
We suggest that, {if other countries insist on monitoring their academic researchers, it would be prudent for them}  to also move towards peer-review-based exercises and away from metrics.

\vspace{-4mm}

\ukrainianpart

\title{Наукометрія з точки зору науковців: не все, що має значення, може бути порахованим}

\author{Р.~Кенна\refaddr{ad1,ad4}, О.~Мриглод\refaddr{ad2}, Б.~Берш \refaddr{ad3,ad4}}
\addresses{\addr{ad1}Дослідницький центр прикладної математики, Університет Ковентрі, Великобританія
\addr{ad2}Інститут фізики конденсованих систем Національної академії наук України, \\вул. Свєнціцького, 1, 79011 Львів, Україна
\addr{ad3} Група статистичної фізики, Університет Лотарингії, Нансі, Франція
 \addr{ad4} Докторський коледж статистичної фізики складних систем,
Ляйпціґ-Лотарингія-Львів-Ковентрі  $({\mathbb L}^4)$, Європа
}

\makeukrtitle

\begin{abstract}
До вподоби це комусь чи ні, але спроби аналізувати та моніторити якість академічних досліджень стають все більш поширеними у цілому світі. Оцінювання ефективності відбувається як на індивідуальному рівні, так і здійснюється для наукових груп, факультетів чи цілих університетів. Багато із таких процедур напряму залежать або ж беруть до уваги прості наукометричні індикатори, а їх висновки можуть впливати на кар'єри, фінансування та престиж.
Проте дуже часто бракує розуміння того, що наукометрики є, в кращому випадку, досить грубим інструментом, і їх неправильне використання може привести до неправильних висновків. Замість того, щоб визнавати підйоми чи падіння у рейтингу для індивідумів, відділень чи університетів на основі таких неідеальних метрик, робляться заклики до регулярної ретельної перевірки індикаторів та до покращення доказової бази в цій ділянці.
Перевірка індикаторів покладена на наукову спільноту, особливо на фізиків, експертів у галузі складних систем та наукометрії. У цій роботі зроблено огляд нещодавніх спроб показати, що деякі поширені метрики не можуть слугувати надійним індикатором наукової якості.

\keywords наукометрія, оцінювання наукових досліджень

\end{abstract}

\end{document}